\documentclass[reprint,preprintnumbers,amsmath,amssymb,superscriptaddress,prb]{revtex4-1}

\usepackage{graphicx,color}

\usepackage{dcolumn}

\usepackage{bm}

\usepackage{hyperref}

\begin{document}

\title{Emergent critical phenomenon in spin-1/2 ferromagnetic-leg ladders: Quasi-one-dimensional Bose--Einstein condensate}

\author{Y. Kono}
\email{y.kono@p.s.osakafu-u.ac.jp}
\affiliation{Department of Physical Science, Osaka Prefecture University, Osaka 599-8531, Japan}
\affiliation{Institute for Solid State Physics, University of Tokyo, Kashiwa 277-8581, Japan}
\author{S. Kittaka}
\affiliation{Institute for Solid State Physics, University of Tokyo, Kashiwa 277-8581, Japan}
\author{H. Yamaguchi}
\affiliation{Department of Physical Science, Osaka Prefecture University, Osaka 599-8531, Japan}
\author{Y. Hosokoshi}
\affiliation{Department of Physical Science, Osaka Prefecture University, Osaka 599-8531, Japan}
\author{T. Sakakibara}
\affiliation{Institute for Solid State Physics, University of Tokyo, Kashiwa 277-8581, Japan}

\date{\today}

\begin{abstract}
We examine the magnetic-field-induced criticality of phase boundary near saturation field $H_{\mathrm{c}}$ in the spin-1/2 ferromagnetic (FM)-leg ladder 3-Cl-4-F-V [=3-(3-chloro-4-fluorophenyl)-1,5-diphenylverdazyl], the predominant interactions of which arise from FM chains (strong-leg type). Critical temperatures were precisely determined through dc magnetization, specific heat, and magnetocaloric effect measurements. The criticality of 3-Cl-4-F-V is characterized by a linear phase boundary with respect to $H_{\mathrm{c}}-H$ near $H\,=\,H_{\mathrm{c}}$. This behavior is similar to that of another strong-leg-type FM-leg ladder. The universal critical behavior in these strong-leg-type FM-leg ladders is expected to demonstrate the theoretically predicted quasi-one-dimensional Bose--Einstein condensation. 
\end{abstract}



\maketitle

\section{Introduction}
The notion of Bose--Einstein condensation (BEC)---a macroscopic quantum state---was first predicted by Bose and Einstein almost a century ago~\cite{Bose_Z.Physik1924,Einstein_SitzberKglPreussAkadWiss1925}. 
Experimental realization of a BEC on ultracold atomic gases of alkali metals received the Nobel prize in Physics~\cite{Cornell_Rev.Mod.Phys.2002,Ketterle_Rev.Mod.Phys.2002a}, and substantial research efforts on this notion have been reported over the last two decades~\cite{Bagnato_RomanianRep.Phys.2015}. 

In condensed matter physics, BEC in quantum magnets is an attractive research topic; the macroscopic quantum state can be realized in quantum spin systems with the mapping between lattice gas bosons and three-dimensional (3D) $XY$-like antiferromagnetic (AFM) ordering~\cite{giamarchi2008bose,RevModPhys.86.563}. Experimental tests for the BEC in quantum magnets have relied on the critical exponent $\phi$ of the phase boundary, defined by $T\sim\left|H_{\mathrm{c}}(T)-H_{\mathrm{c}}(0)\right|^{1/\phi}$, which has been predicted to be $\phi\,=\,3/2$ as $T\,\rightarrow\,0$ for the 3D BEC universality class ($H_{\mathrm{c}}$ typically corresponds to a critical magnetic field on a quantum phase transition [QPT] from a spin disordered or fully polarized state to an AFM ordered state)~\cite{PhysRevB.59.11398,PhysRevLett.84.5868}. In recent decades, numerous experiments on QPTs in real quantum magnets have confirmed the 3D BEC exponent~\cite{RevModPhys.86.563}. The spin dimer systems TlCuCl$_{3}$~\cite{JPSJ.77.013701} and BaCuSi$_{2}$O$_{6}$~\cite{PhysRevB.72.100404} are representative.

Beyond the conventional 3D BEC, recent research interest has focused on the aspects of low dimensionality and frustration. 
Frustration in 3D AFM bcc lattices is expected to cause two-dimensional (2D) BEC behavior, known as ``dimensional reduction''~\cite{sebastian2006dimensional,Batista_Phys.Rev.Lett.2007,Rosch_Phys.Rev.B2007,Schmalian_Phys.Rev.B2008,Laflorencie_Phys.Rev.Lett.2009}.
The relationship between a Tomonaga--Luttinger liquid (TLL) as a one-dimensional (1D) quantum critical state and 3D BEC state has often been discussed for quasi-1D spin-gapped systems, e.g., the spin-1/2 AFM spin ladders (Cu$_{7}$H$_{10}$N)$_{2}$CuBr$_{2}$~\cite{PhysRevLett.108.097201,PhysRevLett.111.106404} and (Cu$_{5}$H$_{12}$N)$_{2}$CuBr$_{4}$~\cite{PhysRevLett.101.247202,PhysRevB.83.054407}. In a recent experimental study on another quasi-1D quantum magnet, a direct change in the critical exponent was observed from the TLL regime to the 3D BEC universality class~\cite{PhysRevB.95.020408}. Another fascinating system for the low dimensionality in BEC is quasi-1D ferromagnetic (FM) chains or quasi-2D FM planes coupled with weak AFM interactions~\cite{PhysRevB.75.134421}. In the theory, the power law of the critical temperature near the saturation field $H_{\mathrm{c}}$ in such systems should exhibit a crossover from $\phi\,=\,3/2$ for the 3D BEC universality class to $\phi\simeq1$ for quasi-1D or quasi-2D cases as the magnetic field moves away from $H_{\mathrm{c}}$. However, few experimental tests for the theoretical proposal have been reported for this theoretical proposal, aside from those for the quasi-2D candidate K$_{2}$CuF$_{4}$~\cite{PhysRevB.95.174406}, particularly for the quasi-1D case.

Our recent studies~\cite{PhysRevB.96.104439,Kono_Phys.Rev.B2018} have revealed that spin-1/2 FM-leg ladders synthesized using verdazyl radical molecules~\cite{PhysRevLett.110.157205,doi:10.7566/JPSJ.83.033707,PhysRevB.89.220402,PhysRevB.91.125104} are a powerful tool for investigating BEC in quasi-1D quantum magnets with predominantly FM interactions. Verdazyl-radical-based molecular crystals typically exhibit isotropic (Heisenberg) spin interactions~\cite{doi:10.7566/JPSJ.83.033707}, the Hamiltonian of such FM-leg ladders can be expressed as
\begin{eqnarray}
\mathcal{H}&=&J_{||}\sum_{i,\alpha}\bm{S}_{i,\alpha}\cdot \bm{S}_{i+1,\alpha}+J_{\perp}\sum_{i}\bm{S}_{i,1}\cdot \bm{S}_{i,2},\label{ladder}
\end{eqnarray}
where the interaction along each leg ($\alpha=1,2$) $J_{||}$ is FM ($J_{||}\,<\,0$), and the rung interaction between the legs $J_{\perp}$ is AFM ($J_{\perp}\,>\,0$). Theoretically, this model has a spin gap with a rung singlet state~\cite{PhysRevB.53.R8848,PhysRevB.67.064419,PhysRevB.72.014449}. In a finite magnetic field, the ground state between the lower critical field $H_{c1}$ and the saturation field $H_{c2}$ has been predicted to be a TLL~\cite{PhysRevB.70.014425}. Such characteristics are analogous to the AFM--AFM case ($J_{||}, J_{\perp}\,>\,0$)~\cite{PhysRevB.59.11398}, which is the most well-studied case. Thus, the conventional 3D BEC state is expected to be induced by weak 3D interladder interactions. Among the three synthesized FM-leg ladders, 3-Br-4-F-V [=3-(3-bromo-4-fluorophenyl)-1,5-diphenylverdazyl], which consists of strong-rung type ($|J_{||}/J_{\perp}|\,<\,1$) ladders, has confirmed this conjecture by exhibiting a spin gap and a 3D BEC exponent near both of the critical fields $H_{c1}$ and $H_{c2}$ at low temperatures~\cite{PhysRevB.89.220402,PhysRevB.96.104439}.

Here, we focus on the FM-leg ladder 3-Cl-4-F-V [=3-(3-chloro-4-fluorophenyl)-1,5-diphenylverdazyl], which consists of strong-leg-type ($|J_{||}/J_{\perp}|\,>\,1$) ladders and was reported as the first realization of a FM-leg ladder~\cite{PhysRevLett.110.157205}. This material shows no spin gap and a double phase transition that differs from those of the other FM-leg ladders. These characteristics may be attributed to frustrated interladder interactions, as predicted by \emph{ab initio} molecular-orbital calculations for the synthesized FM-leg ladders~\cite{PhysRevLett.110.157205,PhysRevB.89.220402,PhysRevB.91.125104}. As the predominant interactions of the strong-leg-type FM-leg ladders consist of FM chains, the quasi-1D BEC behavior of the phase boundary described above is expected to be realized. In fact, another strong-leg-type FM-leg ladder, 3-I-V [=3-(3-iodophenyl)-1,5-diphenylverdazyl], has shown the $\phi=1$ behavior of the 3D ordering phase boundary near the saturation field~\cite{PhysRevB.91.125104,Kono_Phys.Rev.B2018}, similar to the quasi-1D BEC predicted in Ref.~\cite{PhysRevB.75.134421}. Therefore, studying the criticality near the saturation field in 3-Cl-4-F-V with respect to the universality associated with BEC among FM-leg ladders is worthwhile. 

\begin{figure}[b]
\begin{center}
\includegraphics[width=0.75\linewidth, bb=12 14 203 276]{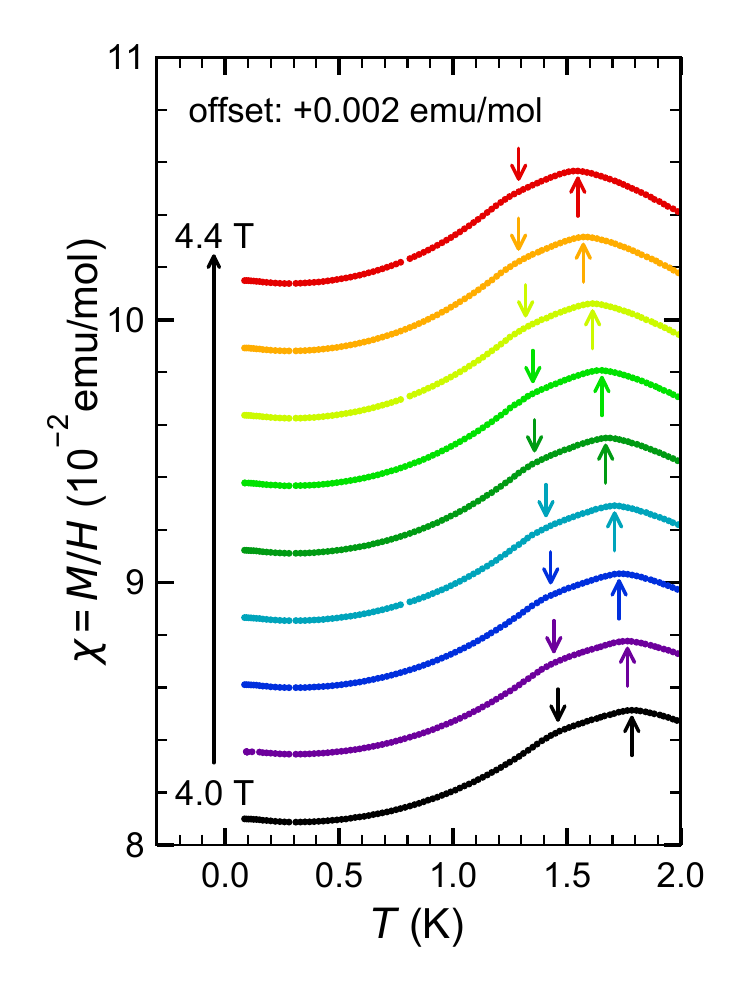}
\caption{Temperature dependence of the magnetic susceptibility $\chi\,=\,M/H$ for magnetic fields ranging from 4.0\,T to 4.4\,T in 0.05\,T steps. Each curve is shifted by $+0.002$\,emu/mol for clarity. The arrows pointing up (down) mark the cusp (kink) anomalies, indicating the upper (lower) critical temperatures (see text).}\label{f1}
\end{center}%
\end{figure}

\begin{figure}[b]
\begin{center}
\includegraphics[width=0.75\linewidth, bb=12 14 203 276]{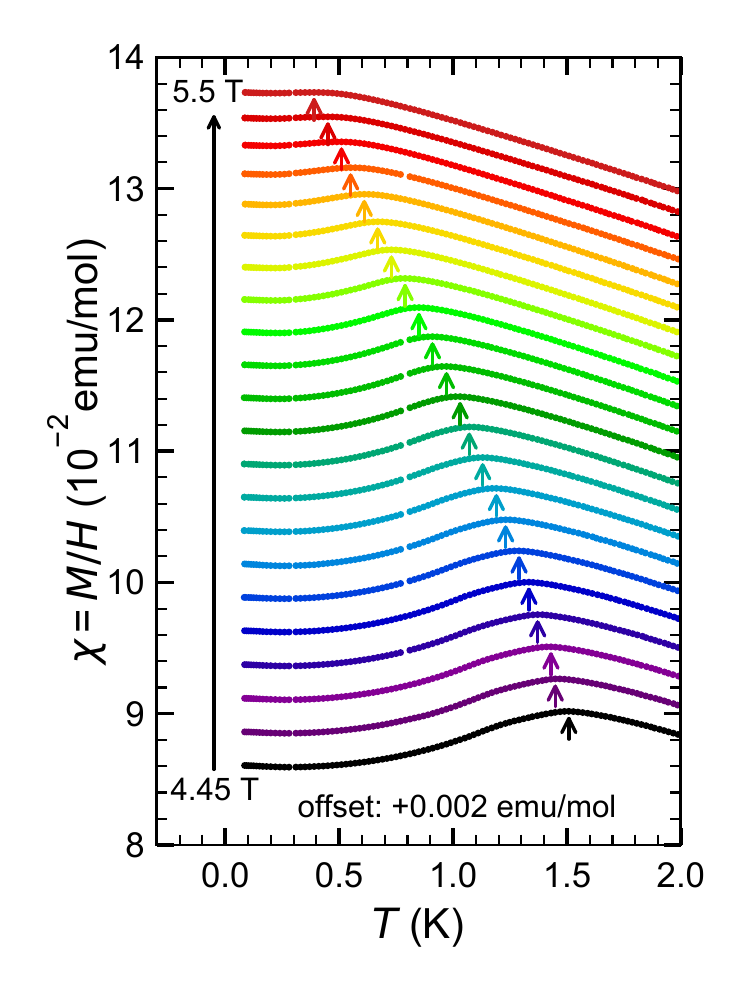}
\caption{Temperature dependence of the magnetic susceptibility $\chi$ for magnetic fields ranging from 4.45\,T to 5.5\,T in 0.05\,T steps. Each curve is shifted by $+0.002$\,emu/mol for clarity. The arrows show the cusp maximum, indicating the critical temperatures (see text).}\label{f2}
\end{center}%
\end{figure}

This study aims to determine the magnetic-field-induced criticality of the phase boundary near saturation in 3-Cl-4-F-V. 
The details of the phase boundary near the saturation field $\mu_{0}H_{\mathrm{c}}\sim$\,5.9\,T were precisely defined through dc magnetization, specific heat, and magnetocaloric effect (MCE) measurements. These measurements reveal that the double phase transition survives below $\sim$4.9\,T, and a single phase boundary region exists from that field to $H_{\mathrm{c}}$. The criticality of the single phase boundary region is characterized by the $\phi=1$ behavior. The universal critical behavior near saturation in the strong-leg-type FM-leg ladders is explored in the context of the theoretically predicted quasi-1D BEC~\cite{PhysRevB.75.134421}.

\section{Experimental}
A single-crystal sample of 3-Cl-4-F-V was prepared as described in our previous report~\cite{PhysRevLett.110.157205}. Dc magnetization measurements were performed using a Faraday-force magnetometer~\cite{JJAP.33.5067}. Specific heat measurements were conducted using the standard quasi-adiabatic heat-pulse method. In these measurements, the same 2.44-mg single crystal was used. The MCE was obtained under ``quasi-adiabatic'' conditions for each 20 mT magnetic field sweep (see Sec.~\ref{sec:mce} for details) using part of the single crystal (0.60\,mg). For all measurements, magnetic fields were applied perpendicular to the $\bm{a}$ axis (perpendicular to the leg direction). As already discussed in the previous report~\cite{PhysRevLett.110.157205}, the magnetic anisotropy is negligibly weak. Although the field direction in the present measurements ($H\perp\bm{a}$) is different from the previous report ($H\,||\,\bm{a}$)~\cite{PhysRevLett.110.157205}, the intrinsic nature of the phase boundary is considered not to be affected.

\section{Results and Discussion}
\subsection{Temperature dependence of the magnetization}

Figures~\ref{f1} and \ref{f2} show the temperature dependence of the magnetic susceptibility $\chi(T)$ ($\chi\,=\,M/H$) near the saturation. As shown in Fig.~\ref{f1}, two anomalies exist in each curve, a cusp and a kink, which can be associated with the double phase transition, as previously reported~\cite{PhysRevLett.110.157205}. As only a cusp was observed for $\chi(T)$ in the previous report, the lower phase boundary was determined only from the specific heat. The clarity of the kink anomalies in $\chi(T)$ for the present data may be attributed to the high quality of the present sample. The cusp also appears sharper than that in the previous report~\cite{PhysRevLett.110.157205}. This fact supports that the peak indicates a phase transition rather than a crossover to the TLL regime. In the magnetic field range shown in Fig.~\ref{f2}, the details of the lower phase boundary cannot be determined from $\chi(T)$ since the kink anomaly becomes ambiguous. 

\begin{figure}[b]
\begin{center}
\includegraphics[width=0.75\linewidth, bb=23 4 268 318]{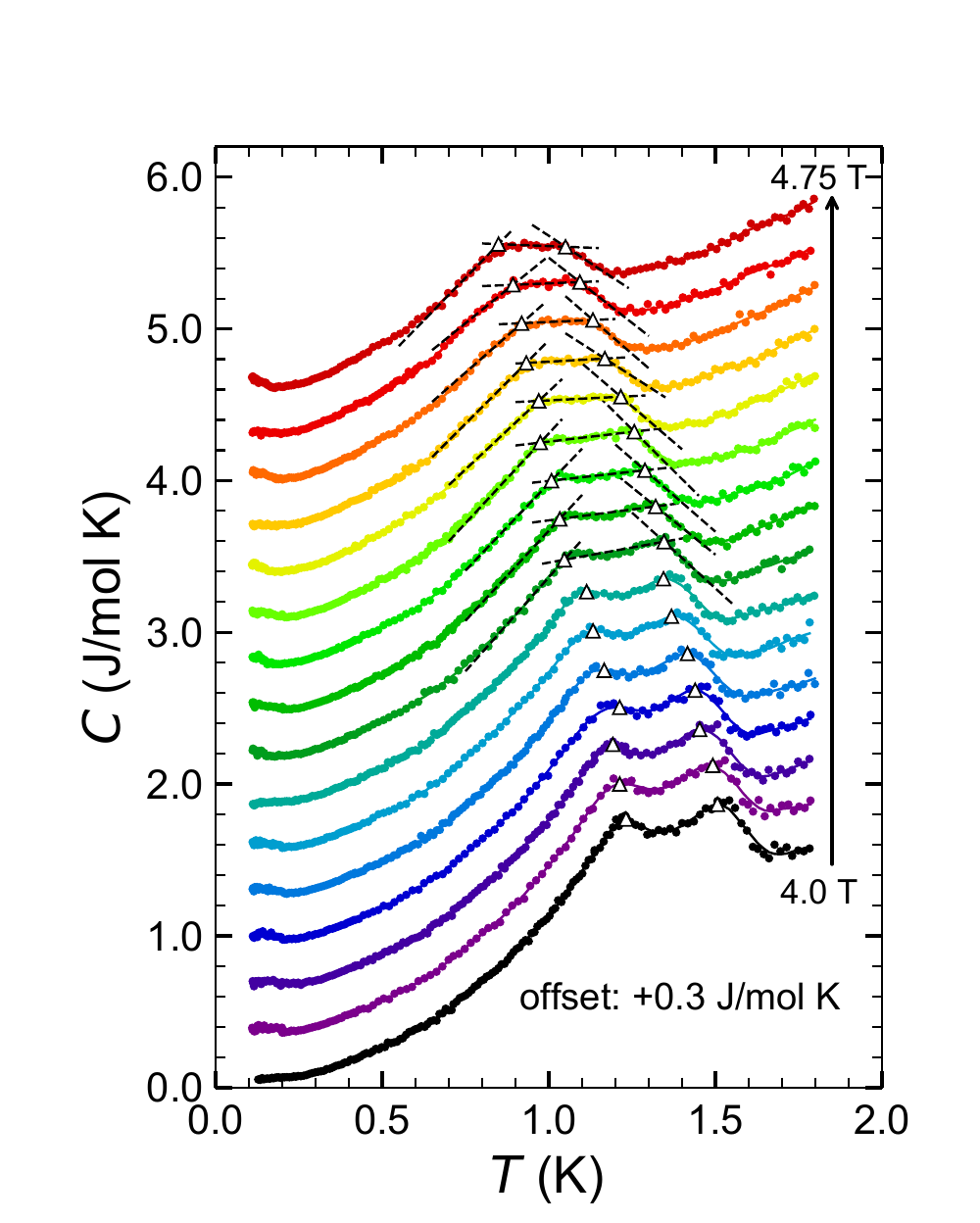}
\caption{Temperature dependence of the specific heat $C$ for magnetic fields ranging from 4.0\,T to 4.75\,T in 0.05\,T steps. Each curve is shifted by $+0.3$\,J/mol K for clarity. The triangles show double-peak structures, indicating the upper and lower critical temperatures. The three dashed lines for each curve above 4.35\,T present linear fits for defining the merged double-peak structures (see text). The squares indicate the onset of the upper anomalies.}\label{f3}
\end{center}%
\end{figure}

\begin{figure}[b]
\begin{center}
\includegraphics[width=0.75\linewidth, bb=23 4 268 318]{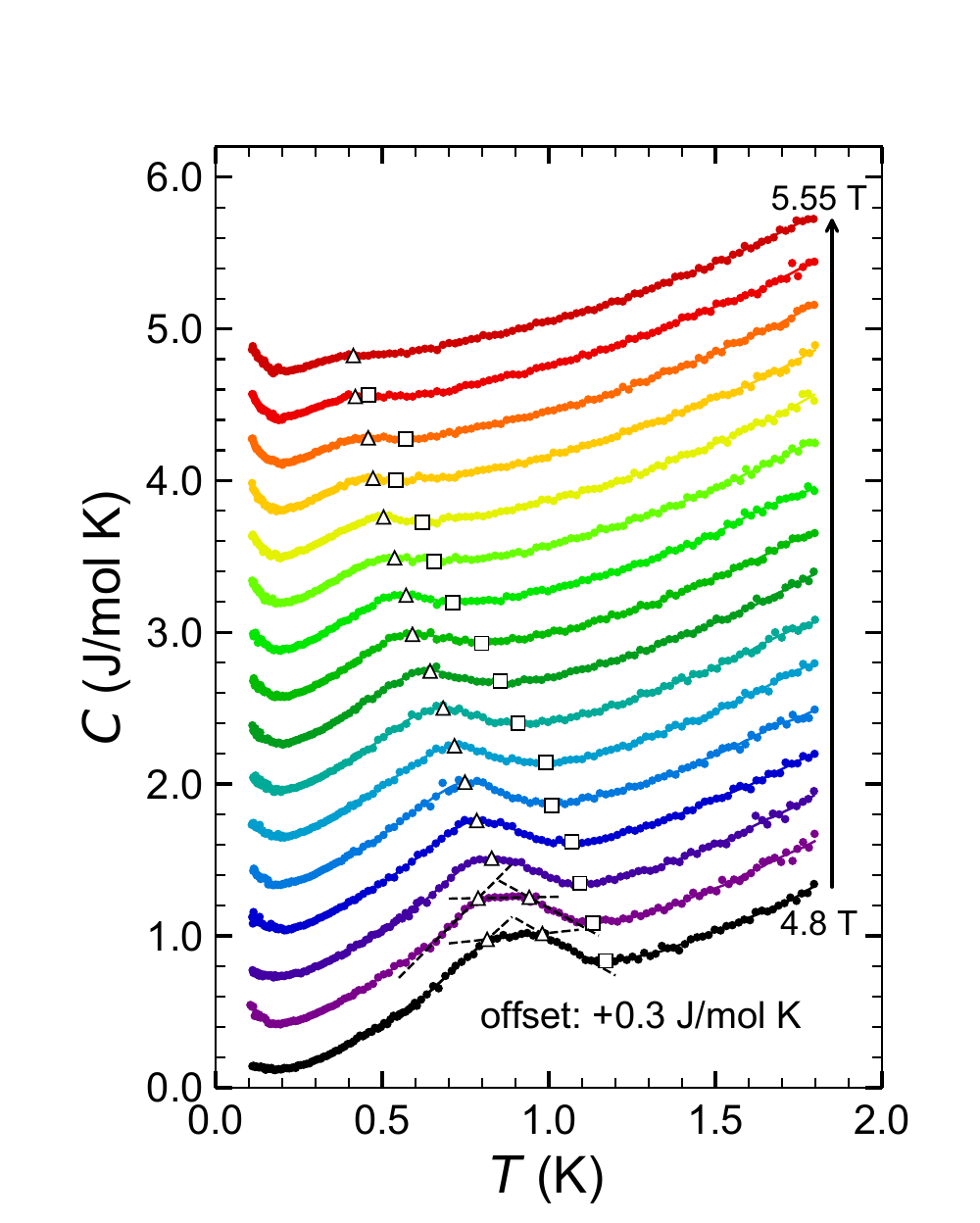}
\caption{Temperature dependence of the specific heat $C$ for magnetic fields ranging from 4.8\,T to 5.55\,T in 0.05\,T steps. Each curve is shifted by $+0.3$\,J/mol K for clarity. The triangles show peak anomalies, indicating critical temperatures. The dashed lines at 4.8 and 4.85\,T are the same as those in Fig.~\ref{f3}. The squares indicate the onset of the upper and single-peak anomalies.}\label{f4}
\end{center}%
\end{figure} 

\subsection{Temperature dependence of the specific heat}\label{sec:sph}

The temperature dependence of the specific heat $C(T)$ is shown in Figs.~\ref{f3} and \ref{f4}. In Fig.~\ref{f3}, $C(T)$ shows obvious double-peak structures, which clearly indicate a double phase transition. The double-peak structures are broader for the high magnetic fields of 4.35--4.85\,T. Eventually, above 4.9\,T, the merged peaks appear to become a single peak (Fig.~\ref{f4})~\footnote{The upturn near the lowest temperature, which is enhanced above 4.8\,T, is attributed to nuclear Schottky contributions primarily arising from $^{1}$H, $^{14}$N, and, $^{35}$Cl.}, implying that only a single phase boundary exists near saturation. To examine the critical behavior of the phase boundary near the saturation, the onset temperature of the peak, as indicated in Figs.~\ref{f3} and \ref{f4}, was adopted as the critical temperature for the upper phase boundary because this definition is in good agreement with that from the $\chi(T)$ and MCE measurements (see also Figs.~\ref{f6}, \ref{f7}(a), and Appendix).

\begin{figure}[t]
\begin{center}
\includegraphics[width=0.75\linewidth, bb=4 4 203 258]{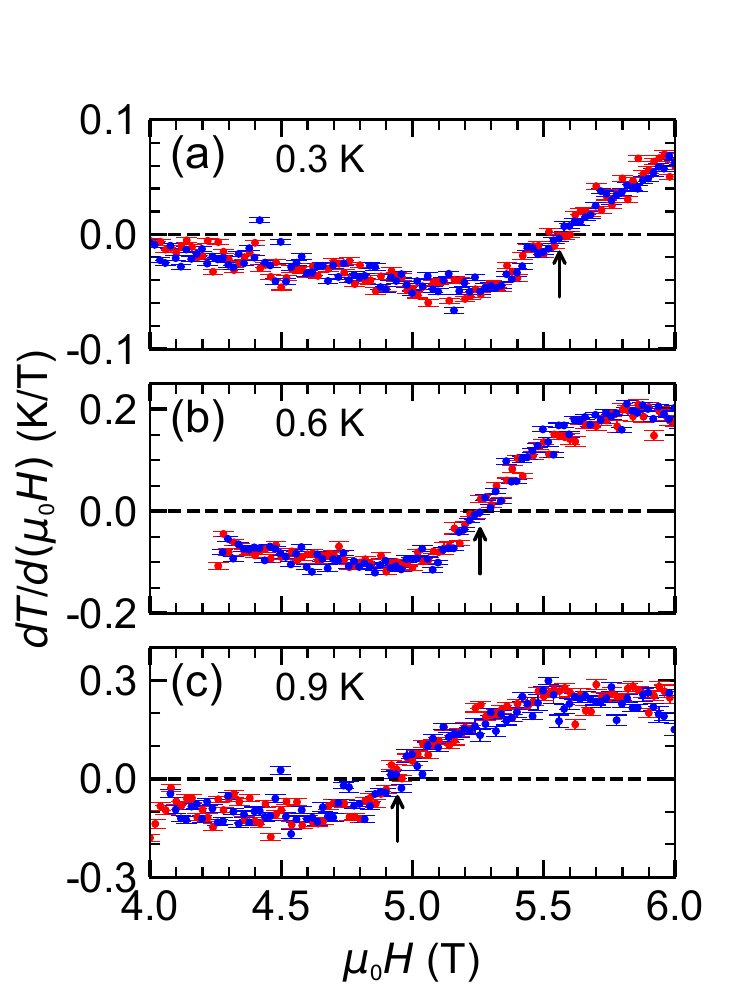}
\caption{MCE $dT/d(\mu_{0}H)$ under ``quasi-adiabatic'' conditions at a fixed bath temperature of (a) 0.3\,K, (b) 0.6\,K, and (c) 0.9\,K. The red and blue symbols correspond to the up and down sweeps, respectively. The arrows show the sign change of $dT/d(\mu_{0}H)$, indicating a phase transition. The error bars denote the fitting errors of the initial slope of $T(H)$ (see text).}\label{f5}
\end{center}%
\end{figure}

\subsection{MCE measurements}\label{sec:mce}

MCE measurements were conducted under ``quasi-adiabatic'' conditions as described below. A general thermodynamic equation for the MCE can be given as the magnetic field derivative of entropy $S$~\cite{Rost_Science2009,Scheven_Phys.Rev.B1997}
\begin{equation}
\left(\frac{\partial S}{\partial H}\right)_{T} = -\frac{\kappa}{T}\frac{\Delta T}{\dot{H}}-\frac{C}{T}\frac{dT}{dH}, \label{eq:mce}
\end{equation}
where $\kappa$ is the thermal conductivity between the sample and the bath, $\Delta T\,=\,T-T_{\mathrm{bath}}$, and $\dot{H}$ is the sweep rate at $\pm$30--40 mT/min for the present measurements. $dT/dH$ was obtained from the initial slope of $T(H)$ immediately after the beginning of each 20 mT magnetic field sweep. After each sweep, the sample temperature $T$ returned to the initial temperature. Consequently, $dT/dH$ values were obtained in 20\,mT intervals throughout the measurement range at a fixed bath temperature. Under the initial conditions for each sweep, $\Delta T$ can be approximated as zero. Thus, the first term in the right side of Eq.~(\ref{eq:mce}) can be ignored (weak coupling limit~\cite{Rost_Science2009}). This condition can be validated by the coincidence of $dT/dH$ in the up and down sweeps, as shown in Fig.~\ref{f5}, because the left term $(\partial S/\partial H)_{T}$ in Eq.~(\ref{eq:mce}) does not change its sign but $\dot{H}$ does with the reversal of the magnetic field~\cite{Kittaka_J.Phys.Soc.Jpn.2018}. A sign change in $dT/dH$ (or an extremum of $T(H)$) in the adiabatic limit indicates a second-order phase transition in a quantum magnet characterized by the BEC state~\cite{RevModPhys.86.563}. As indicated in Fig.~\ref{f5}, the critical magnetic field can be defined from the sign change at a fixed bath temperature because the weak coupling limit is approximated as the adiabatic conditions. 

In Figs.~\ref{f5}(a) and (b), the $dT/dH$ curves show a distinct sign change between 4\,T and 6\,T, which also supports the existence of a single phase boundary in this range, as discussed for the specific heat measurements. As shown in Fig.~\ref{f5}(c), 0.9\,K is within the temperature range at which the double phase transition is expected to exist, as shown in the double-peak structures for $C(T)$. However, whether the sign change exhibits this structure is difficult to determine. The magnetic field range of phase 1 may be too narrow for the two phase boundaries to be distinguished by the MCE measurements under these conditions. 

\begin{figure}[t]
\begin{center}
\includegraphics[width=0.95\linewidth, bb=13 14 348 276]{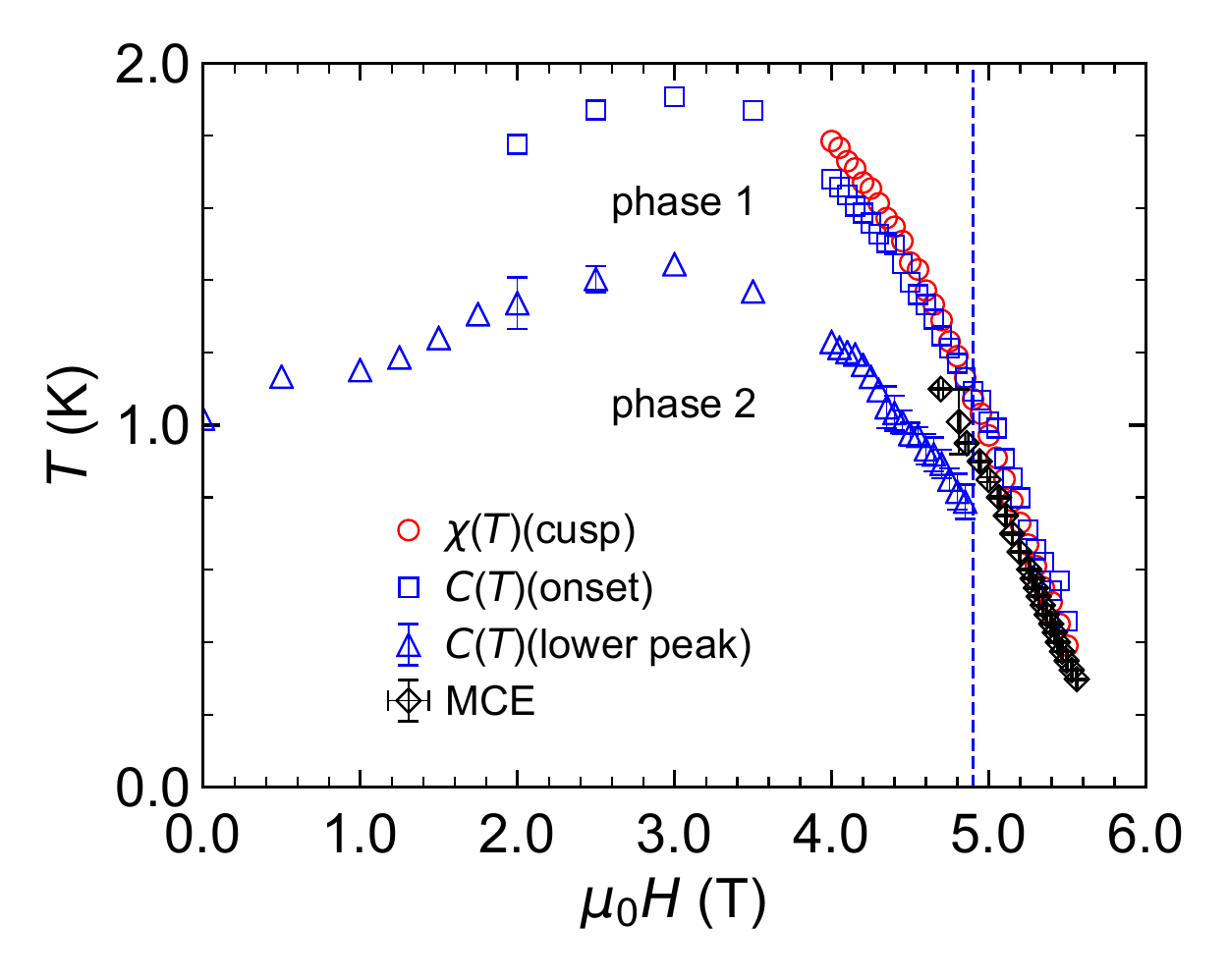}
\caption{$H$--$T$ phase boundaries determined from the present measurements. The circles show the critical temperatures determined from $\chi(T)$ (Figs.~\ref{f1} and \ref{f2}). The squares present temperatures determined the onset of the upper and single-peak anomalies for $C(T)$, and the triangles displays temperatures obtained from the lower anomalies of $C(T)$ (Figs.~\ref{f3} and \ref{f4}). The diamonds present values obtained from the MCE measurements (Fig.~\ref{f5}). The dashed line marks the magnetic field ($\sim$\,4.9\,T) at which the double phase transition appears to merge into a single phase transition.}\label{f6}
\end{center}%
\end{figure}

\subsection{$H$--$T$ phase boundary}\label{pb}

The $H$--$T$ phase boundaries of 3-Cl-4-F-V, defined by the present $\chi(T)$, $C(T)$, and MCE measurements, are summarized in Fig.~\ref{f6}~\footnote{For the upper critical temperature of $C(T)$, the onset temperature of the upper peak, instead of the peak itself, is plotted for the discussion of the criticality in Sec.~\ref{cl}. On the other hand, for the lower critical temperature for $C(T)$, the lower peak temperature below 4.9\,T is plotted for clarity.}. Below 4\,T, the shape of the phase boundaries is almost the same as previously reported phase boundaries~\cite{PhysRevLett.110.157205}. In the phase diagram, the upper phase 1 is closed before the saturation field is reached, i.e., from 4.9\,T to the saturation field, there only exists the phase transition directly from paramagnetic phase to the phase 2. This type of phase diagram structure is analogous to the TLL regime on a 3D BEC phase in quasi-1D spin systems~\cite{PhysRevLett.111.106404,Thielemann_Phys.Rev.B2009,PhysRevB.91.060407,PhysRevB.95.020408}, although in such systems a crossover exists between the paramagnetic phase and the TLL regime. This analogy implies that the phase transition from the paramagnetic state to phase 1 originates from the development of quasi-1D spin correlations, i.e., intraladder correlations. Stripe-FM-like ordering is a probable candidate for phase 1. Theoretical predictions suggest that a small $XY$-like anisotropy in a spin-1/2 FM-leg ladder can induce stripe-FM ordering in a finite magnetic field~\cite{PhysRevB.70.014425}. We can also mention the characteristics of the two phases based on nuclear magnetic resonance (NMR) measurements for the isostructural compound 3-Br-4-F-V~\cite{PhysRevB.89.220402}. The $^{19}$F-NMR spectra for 3-Br-4-F-V revealed that the higher- and lower-temperature phases indicate partial magnetic order and incommensurate long-range order, respectively. Although these compounds are different in the existence of the spin gap, the predominant interladder couplings of 3-Cl-4-F-V evaluated by \emph{ab initio} molecular orbital calculations are the same as those of 3-Br-4-F-V~\cite{doi:10.7566/JPSJ.83.033707}. Thus, the two phases in 3-Cl-4-F-V are also considered to form similar magnetic states. Microscopic measurements may still be required to reveal more detailed magnetic structures of phases 1 and 2.  

\subsection{Universal critical behavior of the phase boundary near saturation}\label{cl}

\begin{figure}[t]
\begin{center}
\includegraphics[width=0.95\linewidth, bb=11 14 390 323]{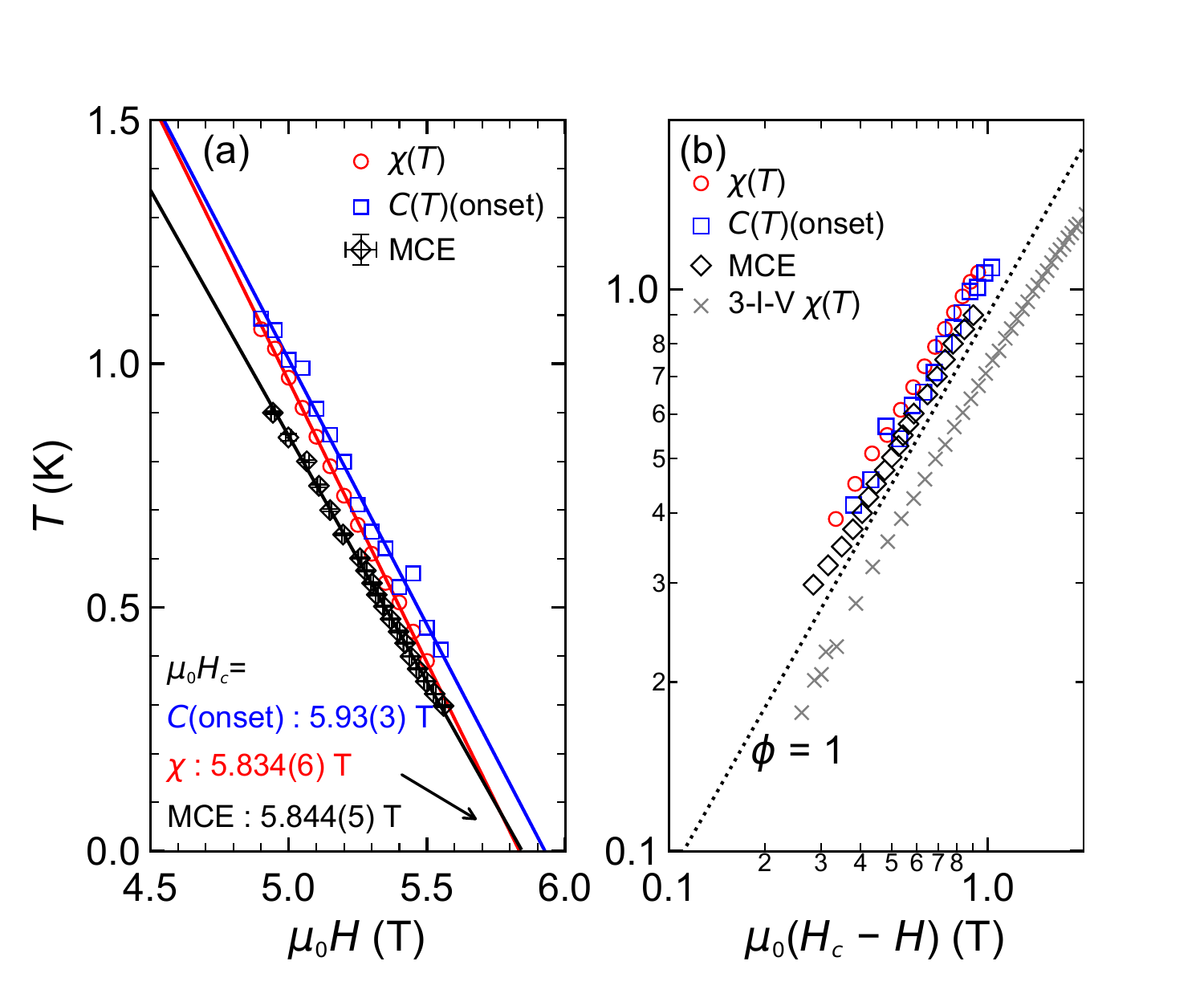}
\caption{(a) Enlarged plot of the $H$--$T$ phase boundary in the single phase boundary region. The solid lines present linear fits for the measurement data above 4.9\,T. (b) Log-log plot of the critical temperature vs. $\mu_{0}(H_{\mathrm{c}}-H)$, where $\mu_{0}H_{\mathrm{c}}$ is estimated from each fitted line in (a). The crosses indicate the critical temperatures obtained from $\chi(T)$ data for another spin-1/2 FM-leg ladder, 3-I-V, as previously reported~\cite{Kono_Phys.Rev.B2018}. The dotted line indicate $T\propto H_{\mathrm{c}}-H$.}\label{f7}
\end{center}%
\end{figure}

Critical phenomena near saturation can reveal the intrinsic nature of the symmetry and dimensionality without a knowledge of the detailed magnetic structures. Figure~\ref{f7}(a) shows an enlarged plot of the phase boundary near saturation. The critical temperature for all measurements appears to be linear with respect to the magnetic field above $\sim$\,4.9\,T, which corresponds to the single phase boundary region indicated in Fig.~\ref{f6}. We extracted the critical field at zero temperature, $H_{\mathrm{c}}$, from the linear fit for each dataset below 1\,K, as plotted in Fig.~\ref{f7}(a), which yields $\mu_{0}H_{\mathrm{c}}\,=\,$5.834(6), 5.844(5), and 5.93(3)\,T for the $\chi(T)$, MCE, and $C(T)$ data, respectively. The critical behavior near saturation can be seen in the log-log plot of the critical temperature with respect to the magnetic-field difference $\mu_{0}(H_{\mathrm{c}}-H)$ for each critical field, as shown in Fig.~\ref{f7}(b). The criticality can be characterized by the $\phi\,=\,1$ behavior at phase boundary $T\,\propto\,(H_{\mathrm{c}}-H)^{1/\phi}$ for a wide temperature range below $\sim$\,0.8\,K, which is distinguished from the $\phi=3/2$ critical exponent for the 3D BEC universality class (see also Appendix). The slight deviation of the MCE data from the $\phi\,=\,1$ line near the lowest temperature may be due to some symmetry breaking terms for the BEC universality such as dipole anisotropy or an effect of nuclear Schottky contributions for $C(T)$, which can become valid at sufficiently low temperatures and high magnetic fields.

The $\phi\,=\,1$ criticality can be also observed in another spin-1/2 FM-leg ladder, 3-I-V, which we previously
reported~\cite{Kono_Phys.Rev.B2018}. In Fig.~\ref{f7}(b), the critical temperature determined from the $\chi(T)$ data for 3-I-V is also shown (see Fig. 6(b) in Ref.~\cite{Kono_Phys.Rev.B2018}), which displays the $\phi\,=\,1$ behavior over a temperature range similar to that for 3-Cl-4-F-V. This trend implies that the universal criticality arises from the predominantly FM interactions in the strong-leg-type FM-leg ladder in these materials. This situation is similar to that for the quasi-1D BEC addressed in Ref.~\cite{PhysRevB.75.134421}.

Several potential reasons may explain how these FM-leg ladders satisfy the conditions of quasi-1D BEC criticality discussed in Ref.~\cite{PhysRevB.75.134421}, i.e., elucidating why they can be regarded as spin-1/2 FM chains with sufficiently weak AFM interchain interactions. First, a spin-1/2 FM-leg ladder can be mapped onto a spin-1/2 FM chain with easy-plane anisotropy~\cite{PhysRevB.70.014425}, corresponding to the model Hamiltonian treated in Ref.~\cite{PhysRevB.75.134421}. The effective easy-plane anisotropy can reinforce the saturation field and thus takes advantage of the observation of the $\phi\,=\,1$ region. By contrast, in an isotropic FM chain, sufficiently weak AFM interchain interactions yield a fairly low saturation field. Second, the frustration of the interladder interactions may suppress the effective interladder magnon interactions. This explanation has already been described for the case of 3-I-V~\cite{Kono_Phys.Rev.B2018}. The weakness of interchain magnon interactions is crucial for observing the $\phi\,=\,1$ crossover, as concluded in Ref.~\cite{PhysRevB.75.134421}. 

Other phenomena may be considered causes of the $\phi\,=\,1$ criticality. One possibility is the quasi-2D BEC, but it has already been excluded for 3-I-V~\cite{Kono_Phys.Rev.B2018}. Another possibility is a disorder-induced transition from BEC to Bose glass~\cite{Zhang_Phys.Rev.B1993,Yamada_Phys.Rev.B2011,Yu_Nature2012,Yao_Phys.Rev.Lett.2014}. However, in the case of the FM-leg ladders, magnetic and non-magnetic impurities are less than only 1\%~\cite{doi:10.7566/JPSJ.83.033707}, which will not affect the intrinsic nature of the magnetic ordering~\footnote{Another possibility is bond disorder, but we did not find any indications of bond disorder in single-crystal x-ray diffraction measurements of this material series~\cite{doi:10.7566/JPSJ.83.033707}. Moreover, we can estimate the degree of bond disorder in other verdazyl-radical-based materials with bond randomness~\cite{Yamaguchi_ScientificReports2017}.}\nocite{Yamaguchi_ScientificReports2017}. Therefore, the $\phi\,=\,1$ criticality in the strong-leg-type FM-leg ladders is promising for the realization of quasi-1D BEC. 

\section{Conclusions}
We have performed detailed thermodynamic measurements of the spin-1/2 strong-leg-type FM-leg ladder 3-Cl-4-F-V near the saturation field $H_{\mathrm{c}}$. A double phase transition is observed up to $\sim$4.9\,T, followed by a single phase boundary region for fields up to $\mu_{0}H_{\mathrm{c}}\,\sim\,5.9$\,T, as revealed by the specific heat and MCE measurements. The critical exponent of the phase boundary $T\,\propto\,(H_{\mathrm{c}}-H)^{1/\phi}$ exhibits the $\phi\,=\,1$ behavior over a wide temperature range in the single phase boundary region for all measurements. This behavior is similar to that of another spin-1/2 strong-leg-type FM-leg ladder, 3-I-V, as previously reported~\cite{Kono_Phys.Rev.B2018}. This universality provides substantial evidence that the strong-leg-type FM-leg ladders are promising candidates for quasi-1D BEC in quasi-1D ferromagnets as theoretically predicted~\cite{PhysRevB.75.134421}. 

\section*{Acknowledgments}
This work was supported in part by KAKENHI Grants No. 16J01784, No. 18H05844, No. 19J01004, No. 18H01161, No. 17H04850, and No. 18H01164 from Japan Society for the Promotion of Science (JSPS), and also by Izumi Science and Technology Foundation. Y. K. would appreciate the support from JSPS as a JSPS Research Fellow. The 3-Cl-4-F-V sample was prepared at Osaka Prefecture University. The dc magnetization, specific-heat, and MCE measurements were conducted at the Institute for Solid State Physics, the University of Tokyo.

\begin{figure}[b]
\begin{center}
\includegraphics[width=0.95\linewidth, bb=11 14 390 323]{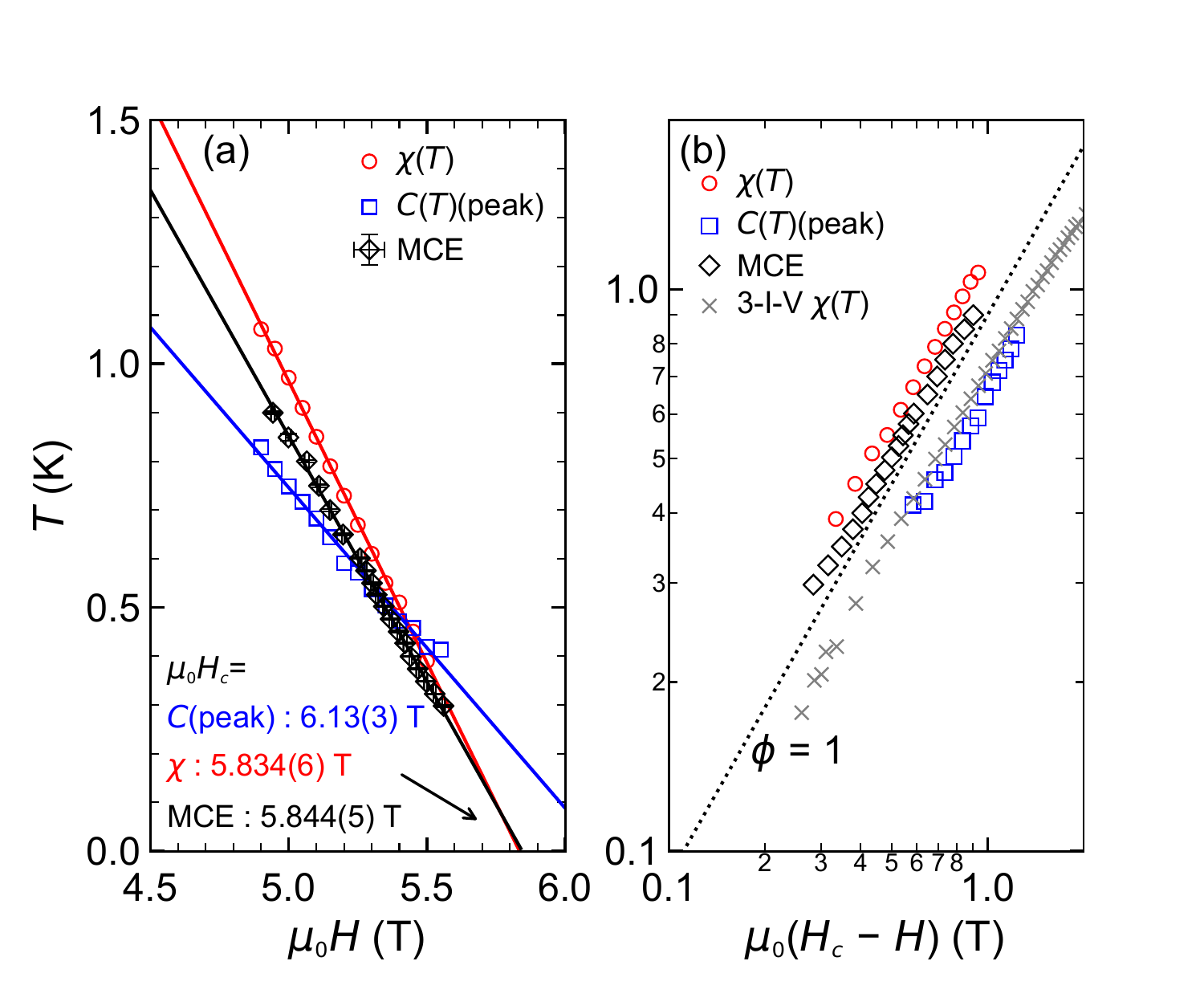}
\caption{(a) Enlarged plot of the $H$--$T$ phase boundary in the single phase boundary region and (b) log-log plot of the critical temperature vs. $\mu_{0}(H_{\mathrm{c}}-H)$, in the same format as Fig.~\ref{f7} for the upper peak temperatures of the specific heat (squares), indicated by the triangles in Fig.~\ref{f4}. }\label{f8}
\end{center}%
\end{figure}

\begin{figure}[t]
\begin{center}
\includegraphics[width=0.75\linewidth, bb=16 5 261 323]{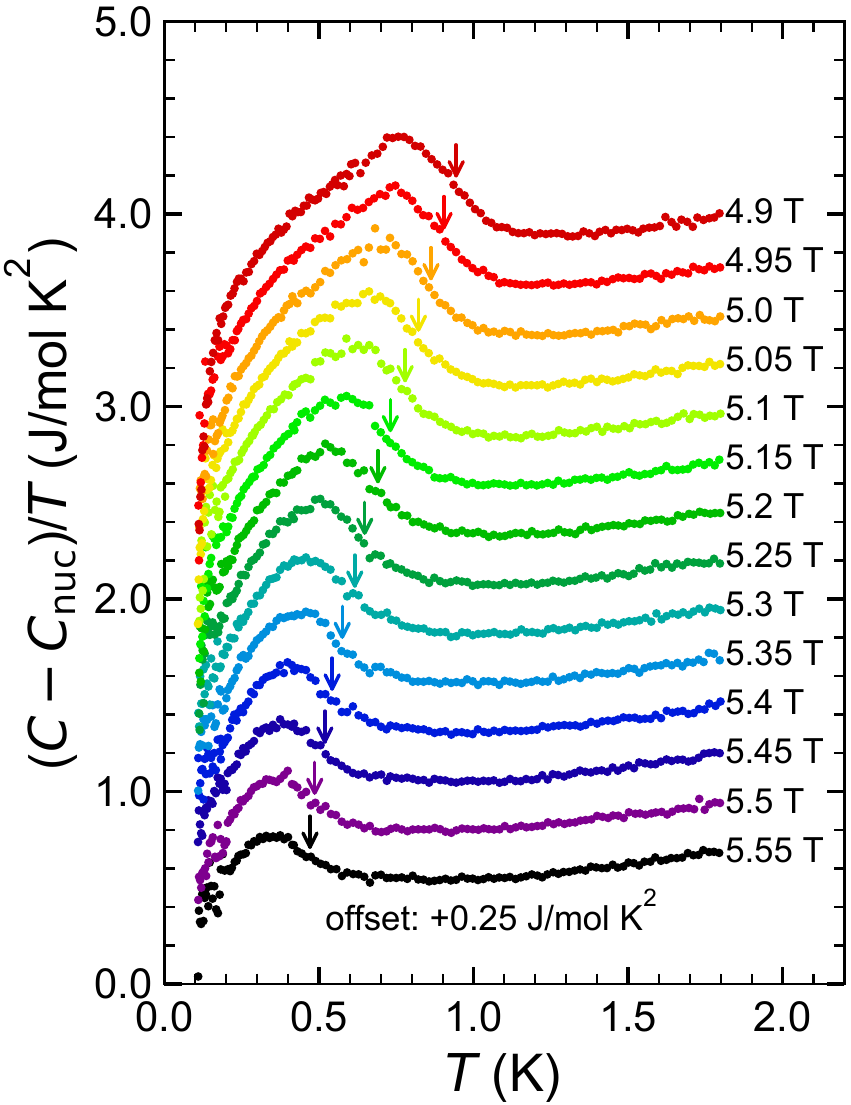}
\caption{Temperature dependence of $(C-C_{\mathrm{nuc}})/T$ for magnetic fields ranging from 4.9\,T to 5.55\,T in 0.05\,T steps, where $C_{\mathrm{nuc}}$ is the estimated nuclear Schottky contributions for each magnetic field. Each curve is shifted by $+0.25$\,J/mol K$^2$ for clarity. The arrows show the midpoints of the two extrema (see text).}\label{f9}
\end{center}%
\end{figure} 

\begin{figure}[t]
\begin{center}
\includegraphics[width=0.95\linewidth, bb=11 14 390 323]{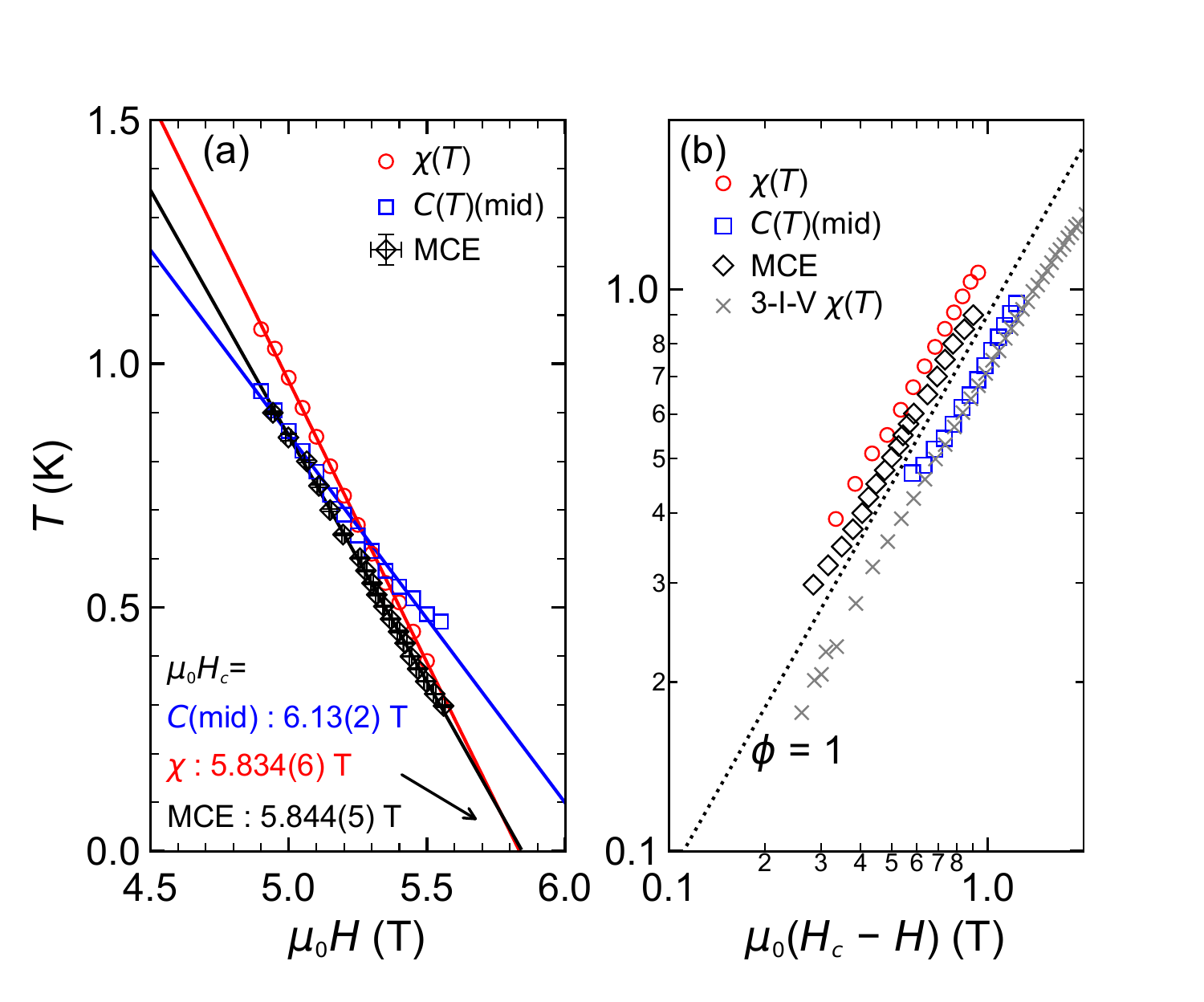}
\caption{(a) Enlarged plot of the $H$--$T$ phase boundary in the single phase boundary region and (b) log-log plot of the critical temperature vs. $\mu_{0}(H_{\mathrm{c}}-H)$, in the same format as Fig.~\ref{f7} for the midpoint temperatures of $(C-C_{\mathrm{nuc}})/T$ (squares), indicated by the arrows in Fig.~\ref{f9}.}\label{f10}
\end{center}%
\end{figure}

\appendix*
\section{Comparison of the definitions of the phase boundary from the specific heat}
Here, we discuss two other definitions of the critical temperature in the specific heat instead of the onset temperature of the peak as shown in Sec.~\ref{sec:sph}. It reveals that the onset temperature is more relevant, and the $\phi\,=\,1$ behavior of the phase boundary is not dependent on the definitions of the critical temperature. 

Simple case of the definition of the critical temperature is the peak temperature as shown in Fig.~\ref{f4} (triangles). Figure~\ref{f8}(a) shows enlarged plot of the $H$--$T$ phase boundary in the single phase boundary region for the peak temperatures of $C(T)$. The slope of the phase boundary defined by the peak temperatures of $C(T)$ obviously deviates from the other measurements. 

One of the best ways to define the critical temperature is to estimate the entropy from numerical integration of $C_{\mathrm{mag}}/T$, where $C_{\mathrm{mag}}$ is the magnetic contributions of the specific heat, and to consider entropy balance on the phase transition. For this purpose, we estimated simple nuclear Schottky contributions for each magnetic field, $C_{\mathrm{nuc}}$, primarily arising from $^{1}$H, $^{14}$N, and, $^{35}$Cl, and subtracted $C_{\mathrm{nuc}}$ from $C(T)$ in Fig.~\ref{f4}. Figure~\ref{f9} shows the temperature dependence of $(C-C_{\mathrm{nuc}})/T$ in the single phase boundary region. Unfortunately, it is difficult to estimate the entropy $S$ because $C_{\mathrm{nuc}}$ is overestimated especially below 0.2\,K. Instead, we estimated the midpoint value of the two extrema, the peak and trough in each curve, and defined the critical temperature which corresponds to the midpoint value (arrows in Fig.~\ref{f9}). 

Figure~\ref{f10}(a) shows enlarged plot of the $H$--$T$ phase boundary in the single phase boundary region for the midpoint temperatures of the $(C-C_{\mathrm{nuc}})/T$ as defined above. The midpoint temperatures are good agreement with the MCE data above about 0.7\,K, but it deviates from the other measurements' data below 0.7\,K. An effect of nuclear Schottky contributions may be a cause of the deviation from the other measurements near the lowest temperature. Thus, considering good agreement of the slope of the phase boundary in the whole single phase boundary region, we adopted the onset temperature for the critical temperature in the main text.      

We also checked the critical behavior for these definitions in the specific heat. Figures.~\ref{f8}(b) and \ref{f10}(b) are log-log plots of the critical temperature vs. $\mu_{0}(H_{\mathrm{c}}-H)$ for the peak and midpoint temperatures, respectively, in the same format as Fig.~\ref{f7}(b). The $\phi\,=\,1$ behavior can also be seen in these plots in the wide temperature range. This fact means that those three defined critical temperatures are also characteristic temperatures for the phase transition and also supports that the $\phi\,=\,1$ behavior is intrinsic nature of 3-Cl-4-F-V.

%

\end{document}